\documentclass[11pt]{article}

\usepackage{blois2008_klinkhamer_preprint,epsfig,xspace} 

\usepackage[latin1]{inputenc}
\usepackage[OT1]{fontenc}

\usepackage{latexsym,amssymb,amsmath,amsfonts}  
\newcommand {\version}{v4}

\newcommand{\citelow}{\cite}                    
\def\Journal#1#2#3#4{{#1} {\bf #2}, #3 (#4)}

\def\NPB{{\em Nucl. Phys.} B}
\def\PLB{{\em Phys. Lett.}  B}
\def\PRL{\em Phys. Rev. Lett.}
\def\PRD{{\em Phys. Rev.} D}

\def\be{\begin{equation}}
\def\ee{\end{equation}}
\def\bea{\begin{eqnarray}}
\def\eea{\end{eqnarray}}

\newcommand{\beq}{\begin{equation}}
\newcommand{\eeq}{\end{equation}}
\newcommand{\beqa}{\begin{eqnarray}}
\newcommand{\eeqa}{\end{eqnarray}}

\newcommand{\DS}[1]{$\mathsf{#1}$\xspace}       
\newcommand{\half}{{\textstyle \frac{1}{2}}}    
\newcommand{\dd}{\mathrm{d}}                    
\newcommand{\id}{\mathrm{d}}                    
\newcommand{\ii}{\mathrm{i}}                    
\newcommand{\vecbf}[1]{\boldsymbol{\mathrm{#1}}}
\newcommand{\vk}{\mathbf{k}}

\begin{document}
\noindent arXiv:0807.2147 [hep-ph]  \hfill KA--TP--16--2008  (\version) 
\vspace*{4cm}
\title{UHECR BOUNDS ON LORENTZ VIOLATION IN THE PHOTON SECTOR}

\author{FRANS R. KLINKHAMER}

\address{Institute for Theoretical Physics, University of Karlsruhe (TH),
76128 Karlsruhe, Germany}

\maketitle\abstracts{The aim of this brief review is to present a case study of
how astrophysics data can be used to get bounds on Lorentz-violating parameters.
For this purpose, a particularly simple Lorentz-violating
modification of the Maxwell theory of photons is considered, which
maintains gauge invariance, \DS{CPT}, and renormalization.
With a standard spin--$\half$ Dirac particle minimally coupled to this
nonstandard photon, the resulting modified-quantum-electrodynamics model
involves nineteen dimensionless ``deformation parameters.''
Ten of these parameters lead to birefringence and are already tightly
constrained by astrophysics. New bounds on the remaining nine nonbirefringent
parameters have been obtained from the inferred absence of vacuum Cherenkov
radiation in ultrahigh-energy-cosmic-ray (UHECR) events.
The resulting astrophysics bounds improve considerably upon current
laboratory bounds and the implications of this ``null experiment''
may be profound, both for elementary particle physics and cosmology.}

\section{Introduction}
\label{FRK-sec:introduction}

One of the fundamental questions of physics is the following:
\emph{does space remain smooth as one probes smaller and smaller distances?}

A conservative limit on the typical length
scale $\ell$ of any nontrivial small-scale structure of space
comes from particle-collider experiments which
see no need to change Minkowski spacetime:
 \beq
\text{LEP/Tevatron:}\quad \ell \lesssim  10^{-18} \;\text{m}
\approx \hbar c/ \big( 200\,\text{GeV} \big) \,. \label{FRK-eq:LEP-bound}
\eeq
Yet, astrophysics (or, more specifically, ``astroparticle physics'')
provides us with very much higher energies.
Here, we intend to present a case study of what astrophysics can do,
provided the underlying physics is well understood.
References will be primarily to research papers.

The proposed case study proceeds in three steps.
First, we discuss the phenomenology of a simple photon-propagation model.
Second, we obtain bounds on the model parameters
from ultrahigh-energy cosmic rays (UHECRs).
Third, we consider the theoretical implications of our results.
The contribution concludes with a brief outlook.

\section{Phenomenology}
\label{FRK-sec:phenomenology}
\subsection{Model}
\label{FRK-sec:model}

Consider the following action for a Lorentz-violating (LV) deformation of
quantum electrodynamics (QED):
\beq\label{FRK-eq:modQED-action}
S_\text{modQED} =
S_\text{modM}+S_\text{standD}\,,
\eeq
with a modified-Maxwell
term~\cite{FRK-ChadhaNielsen1983,FRK-ColladayKostelecky1998},
\begin{subequations}
\beq\label{FRK-eq:modM-action}
S_\text{modM} = \int_{\mathbb{R}^4} \id^4 x \;
\Big( -\textstyle{\frac{1}{4}}\, \big( \eta^{\mu\rho}\eta^{\nu\sigma}
+\kappa^{\mu\nu\rho\sigma}\big)
\,\big( \partial_\mu A_\nu(x) - \partial_\nu A_\mu(x) \big)         
\,\big( \partial_\rho A_\sigma(x) - \partial_\sigma A_\rho(x) \big) 
\Big)\,,
\eeq
and the standard Dirac term for a spin--$\half$ particle with
electric charge $e$ and mass $M$,
\beq\label{FRK-eq:standD-action}
S_\text{standD} =\int_{\mathbb{R}^4} \id^4 x \; \overline\psi(x) \Big(
\gamma^\mu \big(\ii\,\partial_\mu -e A_\mu(x) \big) -M\Big) \psi(x)\,. \eeq
\end{subequations}
Here and in the following, natural units are used with $c=\hbar=1$,
 but, occasionally, $c$ or $\hbar$ are displayed in order to clarify
the physical dimension of a particular expression.
The fundamental constant $c$ now corresponds to the maximum attainable
velocity of the Dirac particle or, more importantly, to the
causal velocity from the underlying Minkowski spacetime with
Cartesian coordinates
$(x^\mu)$ $=$ $(x^0,\boldsymbol{x})$ $=$ $(c\,t,x^1,x^2,x^3)$
and metric $g_{\mu\nu}(x)$ $=$ $\eta_{\mu\nu}$ $\equiv$
$\text{diag}\,(+1$, $-1$, $-1$, $-1)\,$.

The real dimensionless numbers $\kappa^{\mu\nu\rho\sigma}$
in \eqref{FRK-eq:modM-action} are considered to be fixed once and for all,
which makes the model Lorentz noninvariant. However,
the action \eqref{FRK-eq:modQED-action} is still gauge-invariant, \DS{CPT}--even,
and power-counting renormalizable~\cite{FRK-KosteleckyLanePickering2002}.
Clearly, these properties make the model
worthwhile to study theoretically and to verify/falsify experimentally.

In the modified-Maxwell term \eqref{FRK-eq:modM-action},
$\kappa^{\mu\nu\rho\sigma}$ is a
constant background tensor with the same symmetries as the Riemann curvature
tensor and a double trace condition
$\kappa^{\mu\nu}_{\phantom{\mu\nu}\mu\nu}=0$, so that there are $20-1=19$
components.

As ten birefringent parameters are already constrained
at the $10^{-32}$ level~\cite{FRK-KosteleckyMewes2002}
by spectro-polarimetric measurements of distant astronomical sources,
the model can be restricted to the nonbirefringent sector by making the
following \emph{Ansatz}~\cite{FRK-BaileyKostelecky2004}:
\beq\label{FRK-eq:nonbirefringent-ansatz}
\kappa^{\mu\nu\rho\sigma} = \textstyle{\frac{1}{2}} \big(\,
\eta^{\mu\rho}\,\widetilde{\kappa}^{\nu\sigma} -
\eta^{\mu\sigma}\,\widetilde{\kappa}^{\nu\rho} +
\eta^{\nu\sigma}\,\widetilde{\kappa}^{\mu\rho} -
\eta^{\nu\rho}\,\widetilde{\kappa}^{\mu\sigma}  \,\big) , \eeq
for a symmetric and traceless matrix $\widetilde{\kappa}^{\mu\nu}$
with $10-1=9$ components.

Hence, there are nine LV deformation parameters
$\widetilde{\kappa}^{\mu\nu}$ to investigate.
It turns out to be useful to rewrite these parameters as follows:
\newcommand{\third}{\textstyle{\frac{1}{3}}}
\beq\label{FRK-eq:widetilde-kappa-mu-nu-Ansatz}
\big(\widetilde{\kappa}^{\mu\nu}\big) \equiv
\text{diag}\big(1,\third,\third,\third\big)\, \widetilde{\kappa}^{00}
+\big(\delta\widetilde{\kappa}^{\mu\nu}\big),\;\;
\delta\widetilde{\kappa}^{00}=0,
\eeq
with a single
independent parameter $\widetilde{\kappa}^{00}$ for the spatially isotropic
part of $\widetilde{\kappa}^{\mu\nu}$ and eight independent parameters
$\delta\widetilde{\kappa}^{\mu\nu}$ for the nonisotropic part.
Finally, we can express these parameters in terms of the so-called
standard-model-extension (SME) parameters~\cite{FRK-KosteleckyMewes2002}:
\vspace*{0mm} \beq\label{FRK-eq:SME-parameters}
 \left(
  \begin{array}{c}
    \widetilde{\kappa}^{00}\\
    \delta\widetilde{\kappa}^{01} \\
    \delta\widetilde{\kappa}^{02} \\
    \delta\widetilde{\kappa}^{03} \\
    \delta\widetilde{\kappa}^{11} \\
    \delta\widetilde{\kappa}^{12} \\
    \delta\widetilde{\kappa}^{13} \\
    \delta\widetilde{\kappa}^{22} \\
    \delta\widetilde{\kappa}^{23} \\
  \end{array}
\right)
 \equiv
\left(
  \begin{array}{c}
    (3/2)\,\widetilde{\kappa}_\text{tr}\\
    -(\widetilde{\kappa}_{\text{o}+})^{(23)}\\
    -(\widetilde{\kappa}_{\text{o}+})^{(31)}\\
    -(\widetilde{\kappa}_{\text{o}+})^{(12)}\\
    -(\widetilde{\kappa}_{\text{e}-})^{(11)}\\
    -(\widetilde{\kappa}_{\text{e}-})^{(12)}\\
    -(\widetilde{\kappa}_{\text{e}-})^{(13)}\\
    -(\widetilde{\kappa}_{\text{e}-})^{(22)}\\
    -(\widetilde{\kappa}_{\text{e}-})^{(23)}\\
  \end{array}
\right),
 \eeq
where three parity-odd parameters determine an antisymmetric traceless
$3\times 3$ matrix $(\widetilde{\kappa}_{\text{o}+})^{mn}$
and five parity-even parameters a symmetric traceless $3\times 3$ matrix
$(\widetilde{\kappa}_{\text{e}-})^{mn}$.

\subsection{Possible spacetime origin}
\label{FRK-sec:possible-spacetime-origin}

We have already mentioned that the modified-Maxwell model has attractive
properties (gauge invariance, \DS{CPT} invariance, and renormalizability).
Still, it would be better if we had a concrete example of how
the modified-Maxwell model might arise from an underlying theory.

\begin{figure*}[t]
\hspace*{1.5cm}
\includegraphics[width=0.35\textwidth]{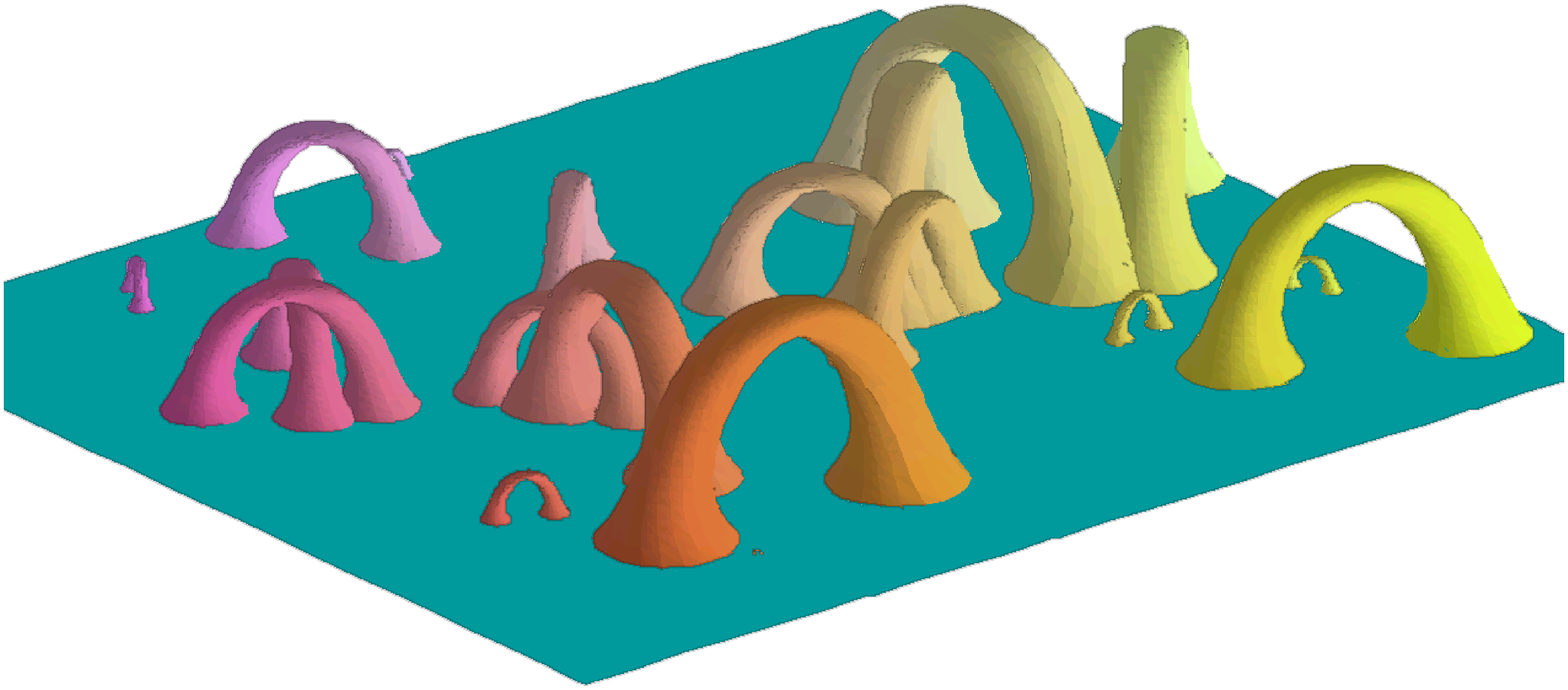}
\hspace*{3.cm}
\includegraphics[width=0.2\textwidth]{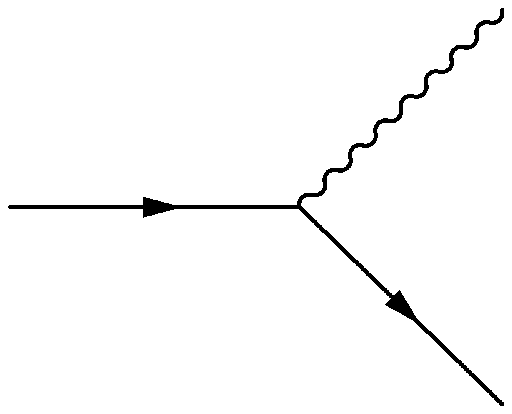}
\caption{Left: classical spacetime-foam manifold.
Right: Feynman diagram for vacuum Cherenkov radiation.}
\label{FRK-fig:foam+Feynman}
\end{figure*}

Precisely this is accomplished by
a calculation~\cite{FRK-BernadotteKlinkhamer2007} of the propagation of
standard photons and standard Dirac particles over a classical spacetime-foam
manifold (Fig.~\ref{FRK-fig:foam+Feynman}--left), which reproduces
a restricted, isotropic version of model \eqref{FRK-eq:modQED-action}:
\beq\label{FRK-eq:alpha0-calculated}
2\,\widetilde{\kappa}_\text{tr}
=
-\widetilde{\sigma}_2\,\widetilde{F} \,,\quad
\delta\widetilde{\kappa}^{\mu\nu}=0\;,
\eeq
in terms of the quadratic coefficient of the modified photon dispersion
relation to be given shortly.

In fact, the simplest possible classical spacetime-foam model has
identical ``defects'' (equal spatial and temporal size $\overline{b}$)
embedded with random orientations in Minkowski spacetime
(equal spatial and temporal average separation $\overline{l}$).
For this type of manifold,
modified dispersion relations for protons ($p$) and photons ($\gamma$)
have been obtained in the large-wavelength
approximation:\begin{subequations}\label{FRK-eq:disprel-general-form}
\beqa \omega^{2}_{p} &\equiv& M_{p}^2\,c_{p}^4/\hbar^2+
c_{p}^2\, k^2 + \text{O}(k^4)\,, \label{FRK-eq:proton-disprel-general-form}
\\[1mm]
\omega^{2}_\gamma &=&
\big(1+ \widetilde{\sigma}_2\,\widetilde{F}\,\big)\;
 c_{p}^2\,k^2 +
\big(\,\widetilde{\sigma}_4\;\widetilde{F}\;\widetilde{b}^{\,2}\,\big)\,
c_{p}^2\, k^4 + \text{O}(k^6)\,, \label{FRK-eq:gamma-disprel-general-form}
\eeqa
\end{subequations}
 with wave number $k \equiv |\vk| \equiv 2\pi/\lambda$,
effective defect on/off factors $\;\widetilde{\sigma}_2,\, \widetilde{\sigma}_4
\in \{\pm 1,\,0\}$, effective defect size $\widetilde{b}$, and effective
defect excluded-volume factor,
\beq\label{FRK-eq:widetildeF}
\widetilde{F} \equiv (\widetilde{b}/\widetilde{l})^4\,,
\eeq
which is assumed to be less than $1$.
Remark that  the coefficient of the quadratic term in
\eqref{FRK-eq:proton-disprel-general-form} is simply defined as $c_{p}^2$
and  that $\widetilde{F}$ appears in both the quadratic and quartic
terms of \eqref{FRK-eq:gamma-disprel-general-form}
[the inclusion of $\widetilde{F}$ in the quartic photon term defines
the meaning of the squared length $\widetilde{b}^{\,2}$].

Calculations of specific space-time-foam
models~\cite{FRK-BernadotteKlinkhamer2007} give the
effective parameters (with tilde) in terms of the
underlying spacetime parameters (with bar):
\beq\label{FRK-eq:parametersfromtau1}
\widetilde{b}
=\beta\; \overline{b}   \,,\quad \widetilde{l} =\lambda\; \overline{l}
\,,\quad \widetilde{\sigma}_2 =-1\,,\quad \widetilde{\sigma}_4 =1\;, \eeq
for positive constants $\beta$ and $\lambda$ of order
unity. From \eqref{FRK-eq:widetildeF} and \eqref{FRK-eq:parametersfromtau1},
the restricted (isotropic)  model
\eqref{FRK-eq:alpha0-calculated} is defined by the fundamental
length scales of the classical space-time-foam model considered, namely,
the typical defect size $\overline{b}$ and the
average separation $\overline{l}$.
[Note that this classical space-time-foam model, in its simplest form,
involves a preferred frame of reference, as the assumed equality of
spatial and temporal defect sizes makes clear.]
In principle, spacetime defects can also generate \DS{CPT}--violating
terms in the effective photon
action~\cite{FRK-KlinkhamerRupp2004,FRK-KlinkhamerRupp2005},
 but the corresponding physical effects are expected to be
suppressed by at least one power
of the fine-structure constant $\alpha \equiv e^2/4\pi \approx 1/137$.

\subsection{Vacuum Cherenkov radiation}
\label{FRK-sec:vacuum-Cherenkov-radiation}

In certain Lorentz-violating photon models, the decay process $p\to p\gamma$
is allowed. This process is similar to that of Cherenkov radiation in material
media~\cite{FRK-Cherenkov1934-37,FRK-FrankTamm1934,FRK-Ginzburg1940,
FRK-Jelley1958,FRK-Afanasiev2004}
 but now occurs already \emph{in vacuo} due to the
modified photon propagation and may, therefore, be called ``vacuum Cherenkov
radiation''~\cite{FRK-Beall1970,FRK-ColemanGlashow1997,FRK-GagnonMoore2004,
FRK-LehnertPotting2004}.
For the modified-QED model \eqref{FRK-eq:modQED-action},
the decay process $p\to p\gamma$ has been studied
classically by Altschul~\cite{FRK-Altschul2007}
and quantum-mechanically (Fig.~\ref{FRK-fig:foam+Feynman}--right) by
Kaufhold and the present author~\cite{FRK-KaufholdKlinkhamer2007}.

The radiated-energy rate of a point particle
with electric charge $Z e$, mass $M> 0$, momentum
$\boldsymbol{q}$, and ultrarelativistic energy $E \sim c\,|\boldsymbol{q}|$
is given by~\cite{FRK-KaufholdKlinkhamer2007}
\beq\label{FRK-eq:dWdt}
\frac{\dd W_\text{modQED}}{\dd t}\, \Bigg|_{E \gg E_\text{thresh}}
 \sim \;\frac{Z^2 e^2}{4\pi}\;\xi(\widehat{\vecbf q}) \; E^2 / \hbar\,,
\eeq
with a direction-dependent coefficient $\xi \geq 0$ and a threshold
energy~\cite{FRK-Altschul2007,FRK-KaufholdKlinkhamer2007}
\beq \label{FRK-eq:Ethreshold}
E_\text{thresh}^2 =
\frac{M^2\,c^4}{R\big[\,2\,\widetilde{\kappa}_\text{tr}+
      2\,\delta\widetilde{\kappa}^{0j}\: \widehat{\vecbf q}^{j}
      +\delta\widetilde{\kappa}^{jk}\:\widehat{\vecbf q}^{j}\,\widehat{\vecbf q}^{k}\big]}
+ \text{O}\left(M^2\, c^4\right),
\eeq
for LV parameters $|\widetilde{\kappa}^{\mu\nu}| \ll 1$ and ramp function
\beq\label{FRK-eq:R}
R[x] \equiv \big(x+|x|\big)/2\,.
\eeq
Note that \eqref{FRK-eq:dWdt} becomes infinite in the classical limit
$\hbar\to 0$, which traces back to the fact that quantum mechanics
provides a frequency cutoff that renders the total radiated energy finite;
see Refs.~\citelow{FRK-Ginzburg1940,FRK-Jelley1958,FRK-Afanasiev2004,
FRK-KaufholdKlinkhamer2007}
for further discussion and references.

Exact tree-level results have been
obtained~\citelow{FRK-KaufholdKlinkhamerSchreck2007,FRK-KlinkhamerSchreck2008}
for the restricted modified-QED model (labeled ``modQED, isotropic case'') with
$\alpha_{0}\equiv (4/3)\,\widetilde{\kappa}^{00}
           \equiv 2\,\widetilde{\kappa}_\text{tr} >0$
and $\delta\widetilde{\kappa}^{\mu\nu}=0$,
which is precisely the model derived in the previous subsection.
Setting again $\hbar=c=1$, the radiated-energy rate for a
spin--$\half$ Dirac point particle (charge $Z e$, mass $M$, and
energy $E=\sqrt{|\vecbf q|^2+M^2}$ above threshold)
is given by
\beqa\label{FRK-eq:dWdt-exact-isotropic}
\hspace*{-13mm}&&
\frac{\dd W_\text{modQED}^\text{isotropic\:case}}{\dd t}\,\Bigg|_{E \geq E_\text{thresh}}
=\;\frac{Z^2 e^2}{4\pi}\;\,\frac{1}{3\,\alpha_0^3 \,E\,\sqrt{E^2-M^2}}\;
\left(\sqrt{\frac{2-\alpha_0}{2+\alpha_0}}\:E-\sqrt{E^2-M^2}\right)^2
\nonumber\\[1mm]
\hspace*{-13mm}
&&\times \Bigg\{\!2\Big(\alpha_0^2+4\alpha_0+6\Big)E^2-\big(2+\alpha_0\big)
\!\left(\!3\big(1+\alpha_0\big)M^2+2\big(3+2\alpha_0\big)\,
\sqrt{\frac{2-\alpha_0}{2+\alpha_0}}\,E\,\sqrt{E^2-M^2}\right)\!\!\Bigg\}.
\eeqa
The high-energy expansion of \eqref{FRK-eq:dWdt-exact-isotropic}
for fixed parameters $\alpha_0$ and $M$ reads
\beqa\label{FRK-eq:dWdt-asymptotic-isotropic}
\hspace*{-14mm}&&
\frac{\dd W_\text{modQED}^\text{isotropic\:case}}{\dd t}\,
\Bigg|_{E \geq E_\text{thresh}}
=\;\frac{Z^2 e^2}{4\pi}\,E^2
\nonumber\\[1mm]
\hspace*{-14mm}&&
\times\Bigg\{
\left(\frac{7}{24}\,\alpha_0-\frac{1}{16}\,\alpha_0^2+\mathsf{O}(\alpha_0^3)\right)
+ \left(-1+\frac{1}{48}\,\alpha_0-\frac{3}{32}\,\alpha_0^2+
\mathsf{O}(\alpha_0^3)\right)\frac{M^2}{E^2}+
\mathsf{O}\left(\frac{M^4}{\alpha_0\,E^4}\right)\Bigg\}\,,
\eeqa
which displays the quadratic behavior of \eqref{FRK-eq:dWdt} for
a constant coefficient $\xi=(7/24)\,\alpha_0+\mathsf{O}(\alpha_0^2)$.
From \eqref{FRK-eq:dWdt-exact-isotropic}, one also obtains the exact
tree-level threshold energy:
\beq\label{FRK-eq:Ethreshold-isotropic}
E^\text{modQED,\,isotropic\:case}_\text{thresh} =
\frac{M}{\sqrt{\alpha_0}}\; \sqrt{1+\alpha_0/2}\:,
\eeq
which reproduces \eqref{FRK-eq:Ethreshold} for a
small positive value of $\alpha_{0}\equiv 2\,\widetilde{\kappa}_\text{tr}$.
Incidentally, the numerical value of $\alpha_{0}$ cannot be too large, as the
factors $\sqrt{2-\alpha_0}$ in \eqref{FRK-eq:dWdt-exact-isotropic} make clear.

As mentioned before, the radiated-energy rates
\eqref{FRK-eq:dWdt} and \eqref{FRK-eq:dWdt-exact-isotropic}
have been derived for point particles.
For realistic particles (protons, nuclei, and photons),
the partonic content can be expected
to significantly modify the prefactors of the radiation rates
 but not the energy thresholds
\eqref{FRK-eq:Ethreshold} and \eqref{FRK-eq:Ethreshold-isotropic},
which follow from energy-momentum conservation
[model \eqref{FRK-eq:modQED-action} violates Lorentz
invariance but maintains spacetime translation invariance]. Note
that calculations of standard classical Cherenkov radiation from different
types of charge distributions give different radiation rates, depending on the
details of the distributions, but unmodified energy thresholds,
at least for the case of static charge distributions in the rest frame and
a nondispersive medium
(cf. Sec.\,7.2 of
Ref.~\citelow{FRK-Afanasiev2004}).\footnote{\label{FRK-ftn:partons}The proton
is a complicated dynamic object (as is the photon)
and ``vacuum Cherenkov radiation''
is not quite the same as standard Cherenkov radiation in a material medium.
For this reason, we have embarked on a
parton calculation~\cite{FRK-KlinkhamerSchreck2008}
of the proton radiated-energy rate
for the isotropic case \eqref{FRK-eq:alpha0-calculated}
with only parameter $\widetilde{\kappa}_\text{tr}$ nonzero
and positive. Preliminary results indicate
that the proton radiated-energy rate is suppressed for
energies  \mbox{$E \in [E_\text{thresh},\,f E_\text{thresh}]$,}
with $E_\text{thresh}$ given by the point-particle
result \eqref{FRK-eq:Ethreshold-isotropic} and $f\approx 1.25$
(based on the results of Table~1 in Ref.~\citelow{FRK-GagnonMoore2004}).}
The bounds on the LV parameters \eqref{FRK-eq:SME-parameters}
derived in the next section rely only on the energy threshold
\eqref{FRK-eq:Ethreshold} and are therefore reliable.

\section{UHECR bounds}
\label{FRK-sec:UHECR-bounds}
\subsection{Cherenkov threshold condition}
\label{FRK-sec:Cherenkov-threshold-condition}

The basic idea~\cite{FRK-Beall1970,FRK-ColemanGlashow1997} consists of three steps:
\vspace*{-0mm}\begin{itemize}
\item
if  vacuum Cherenkov radiation has a threshold energy
$E_\text{thresh}(\widetilde{\kappa})$, then \mbox{UHECRs}
with $E_\text{prim}>E_\text{thresh}$ cannot travel far, as they rapidly
radiate away their energy,
\item\vspace*{-0mm}
this implies that, if an UHECR of energy $E_\text{prim}$ is detected,
its energy must be at or below threshold,
\beq\label{FRK-eq:Cherenkov-condition-general}
E_\text{prim} \leq E_\text{thresh}(\widetilde{\kappa}),
\eeq\item\vspace*{-0mm}
the last inequality gives, using expression \eqref{FRK-eq:Ethreshold},
an upper bound on the
LV parameters~\cite{FRK-KaufholdKlinkhamer2007},
\beq\label{FRK-eq:Cherenkov-condition-modMax}
R\big[\,2\,\widetilde{\kappa}_\text{tr}+
      2\,\delta\widetilde{\kappa}^{0j}\: \widehat{\vecbf q}_\text{prim}^{j}
      +\delta\widetilde{\kappa}^{jk}\:
      \widehat{\vecbf q}_\text{prim}^{j}\,\widehat{\vecbf q}_\text{prim}^{k}\big]
\leq (M_\text{prim}^2\,c^4)/E_\text{prim}^2\,,
\eeq
with primary energy $E_\text{prim}$,
flight direction $\widehat{\vecbf q}_\text{prim}$, and
rest mass $M_\text{prim}$ as input.
\end{itemize}\vspace*{-0mm}
Remark that the Cherenkov threshold condition \eqref{FRK-eq:Cherenkov-condition-modMax}
is only effective for certain combinations of deformation
parameters $\widetilde{\kappa}^{\mu\nu}$ and flight directions
$\widehat{\vecbf q}$, namely, if there is a positive argument of the
ramp function $R$ defined by \eqref{FRK-eq:R}.
Hence, there must be a sufficient number of
UHECR events with primary energies $E_\text{prim}^{(n)}$
and flight directions $\widehat{\vecbf q}_\text{prim}^{(n)}$, in order
to bound \emph{all} deformation parameters \eqref{FRK-eq:SME-parameters}.

\subsection{Bounds on the LV photon parameters $\widetilde{\kappa}^{\mu\nu}$}
\label{FRK-sec:bounds-widetilde-kappa}

The last analysis of
Refs.~\citelow{FRK-KlinkhamerRisse2008a,FRK-KlinkhamerRisse2008b}
is based on the following 29 selected UHECR events:
27 events from Auger--South~\cite{FRK-Abraham-etal2008a},
1 event from Fly's Eye~\cite{FRK-Bird-etal1995}, and
1 event from AGASA~\cite{FRK-Hayashida-etal1994}.
The last two events are taken in order
to provide a better coverage of the northern celestial hemisphere
(later, they can perhaps be replaced by Auger--North events).
The relevant data of these 29 events are given in
Table~\ref{FRK-tab:Auger-events-highE}, where the
uncertainties in the energies are of the order of $25\,\%$
and those in the pointing directions of the order of 1 $\deg$
(see the original references for further discussion).
Adopting a conservative value for the unknown primary mass
(see below), it is this relatively large energy uncertainty which
dominates the error budget of the bounds to be presented
in this subsection and the next.

The 29 primary energies and flight directions
from Table~\ref{FRK-tab:Auger-events-highE}
(flight directions being the opposite of arrival directions)
give 29 inequalities for the deformation parameters
from the Cherenkov threshold condition
\eqref{FRK-eq:Cherenkov-condition-modMax}, where we set
$M_\text{prim}=56\;\text{GeV}/c^2$. [A significant fraction
of these primaries may well be protons~\cite{FRK-Abraham-etal2008a},
 but we prefer to take a conservative value for $M_\text{prim}$.
The used value of $56\;\text{GeV}/c^2$ is even larger than the mass
of an iron nucleus $^{56}\text{Fe}$.]
The 29 inequalities then give the following two--$\sigma$
Cherenkov bounds~\cite{FRK-KlinkhamerRisse2008b} on the nine isolated
SME parameters \eqref{FRK-eq:SME-parameters}
of the nonbirefringent modified-Maxwell model
\eqref{FRK-eq:modQED-action}--\eqref{FRK-eq:nonbirefringent-ansatz}:
\begin{subequations}\label{FRK-eq:SMEbounds-nine}
\begin{eqnarray}
\hspace*{-10mm} (ij)\in \{(23),(31),(12)\} &:&\; \big|
(\widetilde{\kappa}_{\text{o}+})^{(ij)}\big| < 2 \times 10^{-18}\,,
\label{FRK-eq:SMEbounds-nonisotropic-odd}\\[1mm]
\hspace*{-10mm} (kl)\in \{(11),(12),(13),(22),(23)\} &:&\; \big|
(\widetilde{\kappa}_{\text{e}-})^{(kl)}\big| < 4 \times 10^{-18}\,,
\label{FRK-eq:SMEbounds-nonisotropic-even}\\[1mm]
\hspace*{-10mm} &&\;\hspace*{12.5mm} \widetilde{\kappa}_\text{tr} < 1.4
\times 10^{-19}\,, \label{FRK-eq:SMEbounds-isotropic}
\end{eqnarray}
\end{subequations}
for the Sun-centered celestial equatorial coordinate system
in Cartesian coordinates. A single UHECR event suffices
for bound \eqref{FRK-eq:SMEbounds-isotropic} and the $148\;\text{EeV}$
Auger event from Table~\ref{FRK-tab:Auger-events-highE} has been used,
which has a reliable energy calibration.

Based on a $212\;\text{EeV}$ Auger event~\cite{FRK-Abraham-etal2007}
and setting $M_\text{prim}=52\;\text{GeV}/c^2$,
a new result~\cite{FRK-KlinkhamerSchreck2008}
gives the following two-sided bound at the two--$\sigma$ level
on a single universal isotropic parameter:
$- 2\times 10^{-19}<\widetilde{\kappa}_\text{tr}^\text{univ}<6\times 10^{-20}$,
where the lower bound arises solely from partonic effects.
Here, ``universal'' means that the same LV parameter
$\widetilde{\kappa}_\text{tr}$ applies to
all gauge bosons of the Standard Model,
as might be expected from a spacetime-foam model as discussed in
Sec.~\ref{FRK-sec:possible-spacetime-origin}.

\begin{table}[t]
\caption{\label{FRK-tab:Auger-events-highE} Selected UHECR events from Auger
(2004--2007)~\protect\cite{FRK-Abraham-etal2008a},
Fly's Eye (1991)~\protect\cite{FRK-Bird-etal1995},
and AGASA (1993)~\protect\cite{FRK-Hayashida-etal1994}.
Shown are the arrival time (year and Julian day), the primary energy $E$
[in EeV $\equiv 10^{18}$ eV], and
the arrival direction with right ascension $\alpha$ and declination
$\delta$ [both in degrees].}
\vspace{0.1cm}
\renewcommand{\tabcolsep}   {0.90pc}  
\renewcommand{\arraystretch}{0.96}    
\begin{center}
\begin{tabular}{|rrrrr|rrrrr|}
\hline Year & Day &  E$\phantom{2}$  & $\alpha\phantom{.0}$ & $\delta\phantom{.0}$ &
       Year & Day &  E$\phantom{2}$  & $\alpha\phantom{.0}$ & $\delta\phantom{.0}$\\
\hline
 1991 & 288 & 320 &  85.2 &   48.0  & 
\; 2006 &  81 &  79  & 201.1 &  $-$55.3 \\
   1993 & 337 & 210 &  18.9 &   21.1  & 
\; 2006 & 185 &  83  & 350.0 &    9.6 \\
   2004 & 125 &  70  & 267.1 &  $-$11.4 &
\; 2006 & 296 &  69  &  52.8 &   $-$4.5 \\
   2004 & 142 &  84  & 199.7 &  $-$34.9 &
\; 2006 & 299 &  69  & 200.9 &  $-$45.3 \\
   2004 & 282 &  66  & 208.0 &  $-$60.3 &
\; 2007 &  13 & 148  & 192.7 &  $-$21.0 \\
   2004 & 339 &  83  & 268.5 &  $-$61.0 &
\; 2007 &  51 &  58  & 331.7 &    2.9 \\
   2004 & 343 &  63  & 224.5 &  $-$44.2 &
\; 2007 &  69 &  70  & 200.2 &  $-$43.4 \\
   2005 &  54 &  84  &  17.4 &  $-$37.9 &
\; 2007 &  84 &  64  & 143.2 &  $-$18.3 \\
   2005 &  63 &  71  & 331.2 &   $-$1.2 &
\; 2007 & 145 &  78  &  47.7 &  $-$12.8 \\
   2005 &  81 &  58  & 199.1 &  $-$48.6 &
\; 2007 & 186 &  64  & 219.3 &  $-$53.8 \\
   2005 & 295 &  57  & 332.9 &  $-$38.2 &
\; 2007 & 193 &  90  & 325.5 &  $-$33.5 \\
   2005 & 306 &  59  & 315.3 &   $-$0.3 &
\; 2007 & 221 &  71  & 212.7 &   $-$3.3 \\
   2005 & 306 &  84  & 114.6 &  $-$43.1 &
\; 2007 & 234 &  80  & 185.4 &  $-$27.9 \\
   2006 &  35 &  85  &  53.6 &   $-$7.8 &
\; 2007 & 235 &  69  & 105.9 &  $-$22.9 \\
   2006 &  55 &  59  & 267.7 &  $-$60.7 &
&&&&\\ \hline
\end{tabular}
\end{center}
\end{table}

It is important to realize that the Cherenkov bounds
(\ref{FRK-eq:SMEbounds-nine}abc) only depend on the measured energies and flight
directions of the charged cosmic-ray primaries at the top
of the Earth atmosphere. As noted in
Refs.~\citelow{FRK-ColemanGlashow1997,FRK-GagnonMoore2004},
the travel length from vacuum Cherenkov radiation (if operative)
would be of the order of meters rather than megaparsecs. Hence,
we only need to be sure of having observed a charged primary traveling
over a distance of a kilometer, say, in order to apply
the Cherenkov threshold condition \eqref{FRK-eq:Cherenkov-condition-modMax}.
[The previous discussion applies, strictly speaking,
only to event energies above
$1.25\, E_\text{thresh}$ (as mentioned in Ftn.~\ref{FRK-ftn:partons}),
whereas cosmic rays with energies barely above $E_\text{thresh}$
would still have to travel over astronomical distances.]

For comparison, the current laboratory bounds are as follows
(with selected references):
\vspace*{0mm}\begin{itemize}
\item
direct bounds~\cite{FRK-Stanwix-etal2006,FRK-Mueller-etal2007}
on the three parity-odd
nonisotropic parameters in $\widetilde{\kappa}_{\text{o}+}\,$
at the $10^{-12}$ level;
\item\vspace*{0mm}
direct bounds~\cite{FRK-Stanwix-etal2006,FRK-Mueller-etal2007}
on the five parity-even nonisotropic parameters in
$\widetilde{\kappa}_{\text{e}-}\,$ at the $10^{-14}$ to $10^{-16}$ levels;
\item\vspace*{0mm}
direct bound~\cite{FRK-Saathoff-etal2003,FRK-Reinhardt-etal2007}
on the single parity-even isotropic
parameter $\widetilde{\kappa}_\text{tr}\,$ at the $10^{-7}$ level;
\item\vspace*{0mm}
indirect bound~\cite{FRK-Carone-etal2006} on $\widetilde{\kappa}_\text{tr}\,$
at the $10^{-8}$ level from the measured
value~\cite{FRK-Odom-etal2006,FRK-Gabrielse-etal2006} of the electron
anomalous magnetic moment $a_{e}\equiv (g_{e}-2)/2\,$.
\end{itemize}\vspace*{0mm}
Interestingly, the UHECR Cherenkov bounds \eqref{FRK-eq:SMEbounds-nine}
are the strongest where the laboratory bounds are the weakest.
A case in point is the isotropic
parameter $\widetilde{\kappa}_\text{tr}$, which is difficult
to constrain by laboratory experiments (having no sidereal variations)
 but easy to constrain by a single UHECR event (the flight direction being
irrelevant).

The ``leverage factor'' for the UHECR Cherenkov bounds is,
according to \eqref{FRK-eq:Cherenkov-condition-modMax}, given by
$(E_\text{prim}/M_\text{prim}c^2)^2$ and is of the order of
$10^{18}$ for $M_\text{prim}\sim 50\;\text{GeV}/c^2$ and
$E_\text{prim}\sim 50\;\text{EeV}$.
With further information on selected UHECR events
(e.g., the shower-maximum atmospheric depth $X_{\rm max}$ used already
in Ref.~\citelow{FRK-KlinkhamerRisse2008a}),
it may be possible to reduce $M_\text{prim}$ by a factor
10 to $M_\text{prim}\sim 5\;\text{GeV}/c^2$
and, with more and more events becoming available, it may be
possible to select on a typical energy $E_\text{prim}\sim 150\;\text{EeV}$,
thereby increasing the Cherenkov leverage factor by a factor $10^{3}$
to a value of the order of $10^{21}$.
For completeness, the leverage factor for the astrophysics
bounds~\cite{FRK-KosteleckyMewes2002} on the birefringent parameters
is given by $L/\lambda$, which is of the order of $10^{32}$ for a
source distance $L \sim 10^{26}\;\text{m}$ and typical wavelength
$\lambda \sim 10^{-6}\;\text{m}$.
Both leverage factors make clear what the power of astrophysics can be,
provided the physics is well understood.

\subsection{Bounds on the LV photon parameter $\widetilde{b}^{\,2}$}
\label{FRK-sec:bounds-widetilde-b-square}

From the $148\;\text{EeV}$ Auger event in Table~\ref{FRK-tab:Auger-events-highE},
we also get a one--$\sigma$
bound~\cite{FRK-BernadotteKlinkhamer2007} on the general coefficient
of the quartic photon term in \eqref{FRK-eq:gamma-disprel-general-form}:
\beq
\big|\widetilde{\sigma}_4\,\widetilde{F}\, \widetilde{b}^{\,2}\big|
<                                           
\left(2 \times 10^{-35}\,\text{m}\right)^{2}\,,
\label{FRK-eq:Cherenkov-bound-Fbsquare}
\eeq
based on the parton analysis of Ref.~\citelow{FRK-GagnonMoore2004} but
rescaled to $M_\text{prim}=56\;\text{GeV}/c^2$ and
$E_\text{prim}=148\;\text{EeV}$. In fact, the right-hand side
of \eqref{FRK-eq:Cherenkov-bound-Fbsquare} is of order
$(\hbar c^3 M_\text{prim}/E_\text{prim}^2)^2$.
Note that the above bound is two-sided, whereas genuine Cherenkov
radiation would only give a one-sided bound
(the phase velocity of the electromagnetic wave must be less than
the maximum attainable velocity of the charged particle).
The reason for having a two-sided bound
\eqref{FRK-eq:Cherenkov-bound-Fbsquare} is that,
in addition to Cherenkov radiation, another type of process can occur.
The relevant process is proton break-up $p\to p\, e^+\, e^-$ (that is,
pair-production by a virtual gauge boson), which then gives
the ``other'' side of the
bound~\cite{FRK-GagnonMoore2004,FRK-KlinkhamerSchreck2008}.

It may be of interest to mention that the
non-observation of primary photons by the
Pierre Auger Observatory~\cite{FRK-Abraham-etal2007,FRK-Abraham-etal2008b}
has also been used to obtain (modulo some assumptions)
a tight one-sided bound~\cite{FRK-GalaverniSigl2008}
on the quartic term of a modified photon dispersion relation.

In the analysis of Refs.~\citelow{FRK-GagnonMoore2004,FRK-GalaverniSigl2008},
it is taken for granted that the modified photon dispersion relation
corresponds to a consistent theory with, for example,
microcausality and unitarity (cf. the discussion in
Refs.~\citelow{FRK-AdamKlinkhamer2001,FRK-KosteleckyLehnert2001}).
This may very well be the case, provided that the cubic term is absent
from the modified dispersion relation, leaving a quartic term
as the first higher-order term to be considered
(cf. the discussion in Refs.~\citelow{FRK-BernadotteKlinkhamer2007,FRK-Lehnert2003}).
But it is also clear that a real understanding of potential
Lorentz-violating effects can only come from considering a complete
theory, even if it is only an effective theory
such as the one discussed in Sec.~\ref{FRK-sec:possible-spacetime-origin}.

As remarked in an earlier review~\cite{FRK-Klinkhamer2007review},
bound \eqref{FRK-eq:Cherenkov-bound-Fbsquare} disagrees
by many orders of magnitude with a possible
``quantum-gravity'' effect~\cite{FRK-Albert-etal2007-speculation}
in a gamma-ray flare from Mkn
501 as observed by the MAGIC telescope~\cite{FRK-Albert-etal2007-exp}.
Most likely, the observed  time dispersion has an astrophysical origin.
Still, this type of measurement suggests how, in principle,
astrophysical data could give
more than just upper-bounds on the possible small-scale structure of space.

The length scale on the right-hand side of \eqref{FRK-eq:Cherenkov-bound-Fbsquare}
is extraordinarily small, but this inequality becomes less dramatic if
the dimensionless number $\widetilde{F}$ on the left-hand side is also small.
Now this is precisely what has been found in the model calculation
leading up to \eqref{FRK-eq:disprel-general-form},
where the quartic photon coefficient has an extra reduction factor
$\widetilde{F}$ which can be interpreted as the
defect excluded-volume factor \eqref{FRK-eq:widetildeF}
entering the quadratic photon coefficient.
Taking a value $\widetilde{F}=1.5\times 10^{-19}$, just
consistent with the previous result \eqref{FRK-eq:SMEbounds-isotropic},
bound \eqref{FRK-eq:Cherenkov-bound-Fbsquare} becomes
\beq
 \widetilde{b}^{\,2} <         
\big(5 \times 10^{-26} \;\text{m}\big)^2\,,
\label{FRK-eq:Cherenkov-finalbound-b}
\eeq
which remains small compared to the current laboratory
bound \eqref{FRK-eq:LEP-bound}.

\section{Theoretical implications}
\label{FRK-sec:theoretical-implications}

In the previous section, we have established two types of
bounds on Lorentz violation in the photon sector. First, a combined
one--$\sigma$
bound~\cite{FRK-KosteleckyMewes2002,FRK-KlinkhamerRisse2008b}
was obtained on the nineteen Lorentz-violating deformation parameters
of the modified-Maxwell model \eqref{FRK-eq:modQED-action}:
 \beq\label{FRK-eq:kappa-bound}
|\kappa^{\mu\nu\rho\sigma}| < 3 \times 10^{-18} \,,
\eeq
where, for the sake of argument, the
``one-sided'' Cherenkov bound on the isotropic parameter
$\widetilde{\kappa}_\text{tr}$  has also been made ``two-sided''
[as mentioned in the paragraph starting a few lines
under \eqref{FRK-eq:SMEbounds-isotropic},
there is a new two-sided bound at the $10^{-19}$ level].

Second, restricting to the isotropic model
\eqref{FRK-eq:alpha0-calculated}--\eqref{FRK-eq:parametersfromtau1}
with a possible spacetime-foam origin, one--$\sigma$
Cherenkov-type
bounds~\cite{FRK-BernadotteKlinkhamer2007,FRK-GagnonMoore2004,FRK-KlinkhamerSchreck2008}
have been obtained for combinations of the effective
defect size $\widetilde{b}$ and separation $\widetilde{l}$:
\begin{subequations}
\label{FRK-eq:Cherenkov-bounds-concl}
\beqa \widetilde{F} &\equiv&
(\widetilde{b}/\widetilde{l})^4 < 1.5 \times 10^{-19}\,,
\label{FRK-eq:Cherenkov-bounds-concl-F}\\[1mm]
\widetilde{b}\; &<&
  5 \times 10^{-26} \;\text{m}
 \approx \hbar c/\big( 4 \times 10^{9}\,\text{GeV} \big) \,,
\label{FRK-eq:Cherenkov-bounds-concl-widetilde-b} \eeqa
\end{subequations}
where the particular parameters $\widetilde{b}$ and $\widetilde{l}$
are really defined by
the modified dispersion relations (\ref{FRK-eq:disprel-general-form}ab).
For simplicity, we focus our discussion on this
last case with a single quadratic photon coefficient $\widetilde{F}$
and a single quartic photon coefficient $\widetilde{F}\,\widetilde{b}^{\,2}$,
both of which can be interpreted
in terms of a simple spacetime-foam-like structure.

 Bound \eqref{FRK-eq:Cherenkov-bounds-concl-widetilde-b} is
remarkable compared to what can be achieved with particle
accelerators on Earth (recall that the proton beam energy
of the Large Hadron Collider is $7\times 10^{3}\,\text{GeV}$).
As it stands, bound \eqref{FRK-eq:Cherenkov-bounds-concl-widetilde-b}
may be easily satisfied by a quantum-gravity theory with length scale
$l_\text{Planck} \equiv\sqrt{\hbar G/c^3}\approx 1.6 \times 10^{-35}\,\text{m}
\approx  \hbar c/(1.2 \times 10^{19}\,\text{GeV})$.
But for $\text{TeV}$--gravity models~\cite{FRK-ADD1998,FRK-Rubakov2001}
it would be hard to understand why the effective length scale
$\widetilde{b}$ for photon propagation in the three-dimensional world
would be reduced by a numerical factor of the order of $10^6$
compared to the nonperturbative gravity scale
$L_\text{grav} = \hbar c/ E_\text{grav} \sim \hbar c/(4 \,\text{TeV})$.
In fact, the quartic coefficient of the photon dispersion relation
would need an additional factor of the order of $10^{-12}$ or less.
\mbox{\emph{A priori},}
there is no reason to expect the quartic photon coefficient
to be extraordinarily small; see Ref.~\citelow{FRK-BernadotteKlinkhamer2007}
for a heuristic discussion based on so-called Bethe holes~\cite{FRK-Bethe1944}.

Bound \eqref{FRK-eq:Cherenkov-bounds-concl-F}
is even more interesting as it implies that a single-scale
$\big(\widetilde{b}\sim \widetilde{l}\big)$ classical spacetime foam is
ruled out altogether.  This result holds, in fact, for
arbitrarily small values of the defect size
$\widetilde{b}\,$, as long as a classical spacetime makes sense.
In the context of effective theories with Lorentz invariance
violated at an ultraviolet scale $\Lambda$, a similar result holds
that strong LV effects can be expected at low energies~\cite{FRK-Collins-etal2004},
which have not been seen experimentally.

The conclusion is, therefore, that Lorentz invariance remains valid
down towards smaller and smaller distances, which answers in part the
question posed at the beginning of the Introduction.
This conclusion  would hold down to distances at which the classical--quantum
transition occurs, which may happen for distances of the order
of $l_\text{Planck} \approx 10^{-35}\,\text{m}$ or perhaps for distances
given by an entirely new fundamental length scale~\cite{FRK-Klinkhamer2007}.

\section{Outlook\vspace*{-0mm}}
\label{FRK-sec:outlook}

Astrophysics data (in particular, results from UHECRs) show
that a hypothetical quantum spacetime foam must have ``crystalized'' to
a classical spacetime manifold which is remarkably smooth, as
quantified by the defect excluded-volume factor
$(\widetilde{b}/\widetilde{l})^4 \lesssim  10^{-19}  \ll
1$ and Lorentz-violating parameters
$|\kappa^{\mu\nu\rho\sigma}| \lesssim 10^{-18} \ll 1$.
The result applies to any theory of quantum spacetime,
be it Matrix--theory~\cite{FRK-Banks1997,FRK-NishimuraSugino2002} or
loop quantum gravity~\cite{FRK-Rovelli2004}.

The outcome is like having a ``null experiment''
and there is an analogy with the well-known
Michelson--Morley experiment~\cite{FRK-MichelsonMorley1887}:
theory foresees physical effects which are not found by experiment.

This suggests the need for radically new concepts, similar in depth to
the ``relativity of simultaneity'' introduced by Einstein~\cite{FRK-Einstein1905}.
For our problem, a first small step may be the realization that
precisely Lorentz invariance
is crucial at the high-energy frontier and that a new type of
conserved relativistic ``charge'' can play an important role for the
flatness of spacetime by resolving the so-called  cosmological constant
problem~\cite{FRK-KlinkhamerVolovik2008a,FRK-KlinkhamerVolovik2008b,
FRK-KlinkhamerVolovik2008c}.
It remains to be seen if this is a step in the right direction,
 but experiment has, at least, provided theory with a base camp for
the long climb towards the quantum origin of spacetime.

\section*{Acknowledgments\vspace*{-0mm}}
The author thanks his collaborators for their valuable contributions
and \mbox{M. Risse} and G.E. Volovik for useful comments on the manuscript.

\section*{References\vspace*{0mm}}


\begin{thebibliography}{99}

\bibitem{FRK-ChadhaNielsen1983}
S. Chadha and H.B. Nielsen,
\Journal{\NPB}{217}{125}{1983}.

\bibitem{FRK-ColladayKostelecky1998}
D. Colladay and V.A. Kosteleck\'{y},
\Journal{\PRD}{58}{116002}{1998}, arXiv:hep-ph/9809521.

\bibitem{FRK-KosteleckyLanePickering2002}
V.A. Kostelecky, C.D. Lane, and A.G.M. Pickering,
\Journal{\PRD}{65}{056006}{2002}, arXiv:hep-th/0111123.

\bibitem{FRK-KosteleckyMewes2002}
V.A. Kosteleck\'{y} and M. Mewes,
\Journal{\PRD}{66}{056005}{2002}, arXiv:hep-ph/0205211.

\bibitem{FRK-BaileyKostelecky2004}
Q.G. Bailey and V.A. Kosteleck\'{y},
\Journal{\PRD}{70}{076006}{2004}, arXiv:hep-ph/0407252.

\bibitem{FRK-BernadotteKlinkhamer2007}
S. Bernadotte and F.R. Klinkhamer,
\Journal{\PRD}{75}{024028}{2007}, arXiv:hep-ph/0610216.

\bibitem{FRK-KlinkhamerRupp2004}
F.R. Klinkhamer and C. Rupp,
\Journal{\PRD}{70}{045020}{2004}, arXiv:hep-th/0312032.

\bibitem{FRK-KlinkhamerRupp2005}
F.R. Klinkhamer and C. Rupp,
\Journal{\PRD}{72}{017901}{2005}, arXiv:hep-ph/0506071.



\bibitem{FRK-Cherenkov1934-37}
P.A. Cherenkov,
\emph{Dokl. Akad. Nauk Ser. Fiz.} {\bf 2}, 451 (1934);
\emph{Phys. Rev.}  {\bf 52}, 378 (1937).


\bibitem{FRK-FrankTamm1934}
I.M. Frank and I.E. Tamm,
\emph{Dokl. Akad. Nauk Ser. Fiz.} {\bf 14}, 109 (1937).

\bibitem{FRK-Ginzburg1940}
V.L. Ginzburg,
\emph{Zh. Eksp. Teor. Fiz.} {\bf 10}, 589 (1940).

\bibitem{FRK-Jelley1958}
J.V. Jelley,
\emph{\v{C}erenkov Radiation and Its Applications}
(Pergamon Press, London, 1958).

\bibitem{FRK-Afanasiev2004}
G.N. Afanasiev,
\emph{Vavilov--Cherenkov and Synchrotron Radiation: Foundations and Applications}
(Kluwer Academic, Dordrecht, 2004).


\bibitem{FRK-Beall1970}
E.F. Beall,
\Journal{\PRD}{1}{961}{1970}, Sec.~III~A, Cases 1 and 2.

\bibitem{FRK-ColemanGlashow1997}
S.R. Coleman and S.L. Glashow,
\Journal{\PLB}{405}{249}{1997}, arXiv:hep-ph/9703240.

\bibitem{FRK-GagnonMoore2004}
O. Gagnon and G.D. Moore,
\Journal{\PRD}{70}{065002}{2004}, arXiv:hep-ph/0404196.

\bibitem{FRK-LehnertPotting2004}
R. Lehnert and R. Potting,
\Journal{\PRD}{70}{125010}{2004}, arXiv:hep-ph/0408285.

\bibitem{FRK-Altschul2007}
B. Altschul,
\Journal{\PRL}{98}{041603}{2007}, arXiv:hep-th/0609030.

\bibitem{FRK-KaufholdKlinkhamer2007}
C. Kaufhold and F.R. Klinkhamer
\Journal{\PRD}{76}{025024}{2007}, arXiv:0704.3255.

\bibitem{FRK-KaufholdKlinkhamerSchreck2007}
C. Kaufhold, F.R. Klinkhamer, and M. Schreck,
report KA--TP--32--2007 [available from
\texttt{http://www-itp.particle.uni-karlsruhe.de/prep2007.de.shtml}].

\bibitem{FRK-KlinkhamerSchreck2008}
F.R. Klinkhamer and M. Schreck,
to appear in \emph{Phys. Rev.} D, arXiv:0809.3217.

\bibitem{FRK-KlinkhamerRisse2008a}
F.R. Klinkhamer and M. Risse,
\Journal{\PRD}{77}{016002}{2008}, arXiv:0709.2502.

\bibitem{FRK-KlinkhamerRisse2008b}
F.R. Klinkhamer and M. Risse,
\Journal{\PRD}{77}{117901}{2008}, arXiv:0806.4351.

\bibitem{FRK-Abraham-etal2008a}
J. Abraham {\it et al.}  [Pierre Auger Collaboration],
\emph{Astropart. Phys.}  {\bf 29}, 188 (2008),  arXiv:0712.2843.

\bibitem{FRK-Bird-etal1995}
D.J. Bird {\it et al.},
\emph{Astrophys. J.}  {\bf 441}, 144 (1995), arXiv:astro-ph/9410067.

\bibitem{FRK-Hayashida-etal1994}
N. Hayashida {\it et al.},
\Journal{\PRL}{73}{3491}{1994}.

\bibitem{FRK-Abraham-etal2007}
J. Abraham et al.  [Pierre Auger Collaboration],
\emph{Astropart. Phys.}  {\bf 27}, 155 (2007), arXiv:astro-ph/0606619.

\bibitem{FRK-Stanwix-etal2006}
P.L. Stanwix {\it et al.},
\Journal{\PRD}{74}{081101}{2006}, arXiv:gr-qc/0609072.

\bibitem{FRK-Mueller-etal2007}
H. Mueller {\it et al.},
\Journal{\PRL}{99}{050401}{2007}, arXiv:0706.2031.

\bibitem{FRK-Saathoff-etal2003}
G. Saathoff {\it et al.},
\Journal{\PRL}{91}{190403}{2003}.

\bibitem{FRK-Reinhardt-etal2007}
S. Reinhardt {\it et al.},
\emph{Nature Phys.}  {\bf 3}, 861 (2007).

\bibitem{FRK-Carone-etal2006}
C.D. Carone, M. Sher, and M. Vanderhaeghen,
\Journal{\PRD}{74}{077901}{2006}, arXiv:hep-ph/0609150.

\bibitem{FRK-Odom-etal2006}
B. Odom, D. Hanneke, B. D'Urso, and G. Gabrielse,
\Journal{\PRL}{97}{030801}{2006}.

\bibitem{FRK-Gabrielse-etal2006}
G. Gabrielse {\it et al.},
\Journal{\PRL}{97}{030802}{2006}; {\bf 99}, 039902(E) (2007).

\bibitem{FRK-Abraham-etal2008b}
J. Abraham {\it et al.}  [Pierre Auger Collaboration],
\emph{Astropart. Phys.}  {\bf 29}, 243 (2008), arXiv:0712.1147.

\bibitem{FRK-GalaverniSigl2008}
M. Galaverni and G. Sigl,
\Journal{\PRL}{100}{021102}{2008}, arXiv:0708.1737.

\bibitem{FRK-AdamKlinkhamer2001}
C. Adam and F.R. Klinkhamer,
\Journal{\NPB}{607}{247}{2001}, arXiv:hep-ph/0101087.

\bibitem{FRK-KosteleckyLehnert2001}
V.A. Kosteleck\'{y} and R. Lehnert,
\Journal{\PRD}{63}{065008}{2001}, arXiv:hep-th/0012060.

\bibitem{FRK-Lehnert2003}
R. Lehnert,
\Journal{\PRD}{68}{085003}{2003}, arXiv:gr-qc/0304013.

\bibitem{FRK-Klinkhamer2007review}
F.R. Klinkhamer,
\emph{AIP Conf. Proc.}  {\bf 977}, 181 (2008), arXiv:0710.3075.

\bibitem{FRK-Albert-etal2007-speculation}
J. Albert {\it et al.}  [MAGIC Collaboration],
\mbox{\Journal{\PLB}{668}{253}{2008}, arXiv:0708.2889v3.}

\bibitem{FRK-Albert-etal2007-exp}
J. Albert {\it et al.}  [MAGIC Collaboration],
\emph{Astrophys. J.}  {\bf 669}, 862 (2007), arXiv:astro-ph/0702008.

\bibitem{FRK-ADD1998}
N. Arkani--Hamed, S. Dimopoulos, and G.R. Dvali,
\Journal{\PRD}{59}{086004}{1999}, arXiv:hep-ph/9807344.

\bibitem{FRK-Rubakov2001}
V.A. Rubakov,
\emph{Phys. Usp.}  {\bf 44}, 871 (2001), 
arXiv:hep-ph/0104152.

\bibitem{FRK-Bethe1944}
H.A. Bethe,
\emph{Phys. Rev.} {\bf 66}, 163 (1944).

\bibitem{FRK-Collins-etal2004}
J. Collins, A. Perez, D. Sudarsky, L. Urrutia, and H. Vucetich,
\Journal{\PRL}{93}{191301}{2004}, arXiv:gr-qc/0403053;
J. Collins, A. Perez, and D. Sudarsky,
arXiv:hep-th/0603002.

\bibitem{FRK-Klinkhamer2007}
F.R. Klinkhamer,
\emph{JETP Lett.} {\bf 86}, 73 (2007), arXiv:gr-qc/0703009.

\bibitem{FRK-Banks1997}
T. Banks,
\emph{Nucl. Phys. Proc. Suppl.} {\bf 67}, 180 (1998), arXiv:hep-th/9710231.

\bibitem{FRK-NishimuraSugino2002}
J. Nishimura and F. Sugino,
\emph{JHEP} {\bf 0205}, 001 (2002), arXiv:hep-th/0111102.

\bibitem{FRK-Rovelli2004}
C. Rovelli,
\emph{Quantum Gravity}
(Cambridge University Press, Cambridge, England, 2004).

\bibitem{FRK-MichelsonMorley1887}
A.A. Michelson and E.W. Morley,
\emph{Am. J. Sci.}   {\bf 34}, 333 (1887).

\bibitem{FRK-Einstein1905}
A. Einstein,
\emph{Ann. Phys. (Leipzig)} {\bf 17}, 891 (1905)
[reprinted \emph{ibid.} {\bf 14}, S1, 194 (2005)].

\bibitem{FRK-KlinkhamerVolovik2008a}
F.R. Klinkhamer and G.E. Volovik,
\Journal{\PRD}{77}{085015}{2008}, arXiv:0711.3170.

\bibitem{FRK-KlinkhamerVolovik2008b}
F.R. Klinkhamer and G.E. Volovik,
\Journal{\PRD}{78}{063528}{2008}, arXiv:0806.2805.

\bibitem{FRK-KlinkhamerVolovik2008c}
F.R. Klinkhamer and G.E. Volovik,
{\em JETP Lett.} {\bf 88}, 289 (2008), arXiv:0807.3896.  

\end{thebibliography}
\end{document}